\documentclass[12pt,nohyper,notoc]{JHEP}
\usepackage{graphics}
\usepackage{amssymb,epsfig,amsmath,euscript,array,cite}

\newcounter{multieqs}

\newcommand{\bq}{\begin{equation}}
\newcommand{\fq}{\end{equation}}
\newcommand{\bqr}{\begin{eqnarray}}
\newcommand{\fqr}{\end{eqnarray}}

\newcommand{\be}{\begin{equation}}
\newcommand{\ee}{\end{equation}}

\newcommand{\bm}[1]{\mbox{\boldmath $#1$}}

\def\bd{\begin{document}}
\def\ed{\end{document}}
\def\nn{\nonumber}
\def\bea{\begin{eqnarray}}
\def\eea{\end{eqnarray}}
\let\bm=\bibitem
\let\la=\label

\def\npb#1#2#3{Nucl. Phys. {\bf{B#1}} #3 (#2)}
\def\plb#1#2#3{Phys. Lett. {\bf{#1B}} #3 (#2)}
\def\prl#1#2#3{Phys. Rev. Lett. {\bf{#1}} #3 (#2)}
\def\prd#1#2#3{Phys. Rev. {D \bf{#1}} #3 (#2)}
\def\cmp#1#2#3{Comm. Math. Phys. {\bf{#1}} #3 (#2)}
\def\cqg#1#2#3{Class. Quantum Grav. {\bf{#1}} #3 (#2)}
\def\nppsa#1#2#3{Nucl. Phys. B (Proc. Suppl.) {\bf{#1A}}#3 (#2)}
\def\ap#1#2#3{Ann. of Phys. {\bf{#1}} #3 (#2)}
\def\ijmp#1#2#3{Int. J. Mod. Phys. {\bf{A#1}} #3 (#2)}
\def\rmp#1#2#3{Rev. Mod. Phys. {\bf{#1}} #3 (#2)}
\def\mpla#1#2#3{Mod. Phys. Lett. {\bf A#1} #3 (#2)}
\def\jhep#1#2#3{J. High Energy Phys. {\bf #1} #3 (#2)}
\def\atmp#1#2#3{Adv. Theor. Math. Phys. {\bf #1} #3 (#2)}

\newcommand{\EQ}[1]{\begin{equation} #1 \end{equation}}
\newcommand{\AL}[1]{\begin{subequations}\begin{align} #1 \end{align}\end{subequations}}
\newcommand{\SP}[1]{\begin{equation}\begin{split} #1 \end{split}\end{equation}}
\newcommand{\ALAT}[2]{\begin{subequations}\begin{alignat}{#1} #2 \end{alignat}\end{subequations}}
\def\beqa{\begin{eqnarray}} 
\def\eeqa{\end{eqnarray}} 
\def\beq{\begin{equation}} 
\def\eeq{\end{equation}} 

\def\N{{\cal N}}
\def\sst{\scriptscriptstyle}
\def\thetabar{\bar\theta}
\def\Tr{{\rm Tr}}
\def\one{\mbox{1 \kern-.59em {\rm l}}}
\def\uno{\mbox{1 \kern-.59em {\rm l}}} 

\def\a{\alpha}		\def\da{{\dot\alpha}}
\def\b{\beta}		\def\db{{\dot\beta}}
\def\c{\gamma} 	\def\C{\Gamma}	\def\cdt{\dot\gamma}
\def\d{\delta}	\def\D{\Delta}	\def\ddt{\dot\delta}
\def\e{\epsilon}		\def\vare{\varepsilon}
\def\f{\phi}	\def\F{\Phi}	\def\vvf{\f}
\def\h{\eta}
\def\k{\kappa}
\def\l{\lambda}	\def\L{\Lambda}
\def\m{\mu}	\def\n{\nu}
\def\p{\pi}	\def\P{\Pi}
\def\r{\rho}
\def\s{\sigma}	\def\S{\Sigma}
\def\t{\tau}
\def\th{\theta}	\def\Th{\Theta}	\def\vth{\vartheta}
\def\X{\Xeta}
\def\z{\zeta}

\def\cA{{\cal A}} \def\cB{{\cal B}} \def\cC{{\cal C}}
\def\cD{{\cal D}} \def\cE{{\cal E}} \def\cF{{\cal F}}
\def\cG{{\cal G}} \def\cH{{\cal H}} \def\cI{{\cal I}}
\def\cJ{{\cal J}} \def\cK{{\cal K}} \def\cL{{\cal L}}
\def\cM{{\cal M}} \def\cN{{\cal N}} \def\cO{{\cal O}}
\def\cP{{\cal P}} \def\cQ{{\cal Q}} \def\cR{{\cal R}}
\def\cS{{\cal S}} \def\cT{{\cal T}} \def\cU{{\cal U}}
\def\cV{{\cal V}} \def\cW{{\cal W}} \def\cX{{\cal X}}
\def\cY{{\cal Y}} \def\cZ{{\cal Z}}

\def\ua{\underline{\alpha}}
\def\ub{\underline{\phantom{\alpha}}\!\!\!\beta}
\def\uc{\underline{\phantom{\alpha}}\!\!\!\gamma}
\def\um{\underline{\mu}}
\def\ud{\underline\delta}
\def\ue{\underline\epsilon}
\def\una{\underline a}\def\unA{\underline A}
\def\unb{\underline b}\def\unB{\underline B}
\def\unc{\underline c}\def\unC{\underline C}
\def\und{\underline d}\def\unD{\underline D}
\def\une{\underline e}\def\unE{\underline E}
\def\unf{\underline{\phantom{e}}\!\!\!\! f}\def\unF{\underline F}
\def\unm{\underline m}\def\unM{\underline M}
\def\unn{\underline n}\def\unN{\underline N}
\def\unp{\underline{\phantom{a}}\!\!\! p}\def\unP{\underline P}
\def\unq{\underline{\phantom{a}}\!\!\! q}
\def\unQ{\underline{\phantom{A}}\!\!\!\! Q}
\def\unH{\underline{H}}

\def\As {{A \hspace{-6.4pt} \slash}\;}
\def\Ds {{D \hspace{-6.4pt} \slash}\;}
\def\ds {{\del \hspace{-6.4pt} \slash}\;}
\def\ss {{\s \hspace{-6.4pt} \slash}\;}
\def\ks {{ k \hspace{-6.4pt} \slash}\;}
\def\ps {{p \hspace{-6.4pt} \slash}\;}
\def\pas {{{p_1} \hspace{-6.4pt} \slash}\;}
\def\pbs {{{p_2} \hspace{-6.4pt} \slash}\;}

\def\Fh{\hat{F}}
\def\Xh{\hat{X}}
\def\ah{\hat{a}}
\def\xh{\hat{x}}
\def\yh{\hat{y}}
\def\ph{\hat{p}}
\def\xih{\hat{\xi}}

\def\psit{\tilde{\psi}}
\def\Psit{\tilde{\Psi}}
\def\tht{\tilde{\th}}
 
\def\At{\tilde{A}}
\def\Qt{\tilde{Q}}
\def\Rt{\tilde{R}}

\def\ft{\tilde{f}}
\def\pt{\tilde{p}}
\def\qt{\tilde{q}}
\def\vt{\tilde{v}}

\def\delb{\bar{\partial}}
\def\bz{\bar{z}}
\def\Db{\bar{D}}

\def\d{\delta}\def\D{\Delta}\def\ddt{\dot\delta}

\def\pa{\partial} \def\del{\partial}
\def\xx{\times}

\def\trp{^{\top}}
\def\inv{^{-1}}
\def\dag{^{\dagger}}
\def\pr{^{\prime}}

\def\rar{\rightarrow}
\def\lar{\leftarrow}
\def\lrar{\leftrightarrow}

\newcommand{\0}{\,\!}      
\def\one{1\!\!1\,\,}
\def\im{\imath}
\def\jm{\jmath}

\newcommand{\tr}{\mbox{tr}}
\newcommand{\slsh}[1]{/ \!\!\!\! #1}

\def\vac{|0\rangle}
\def\lvac{\langle 0|}

\def\hlf{\frac{1}{2}}
\def\ove#1{\frac{1}{#1}}

\def\Box{\square}
\def\ZZ{\mathbb{Z}}
\def\CC#1{({\bf #1})}
\def\bcomment#1{}
\def\bfhat#1{{\bf \hat{#1}}}
\def\VEV#1{\left\langle #1\right\rangle}

\newcommand{\ex}[1]{{\rm e}^{#1}} \def\ii{{\rm i}}

\title{Noncommutativity and Model Building}

\author{Chong-Sun Chu$^{a}$, Valentin V. Khoze$^b$ 
and Gabriele Travaglini$^b$\\
$^a$Centre for Particle Theory, Department of Mathematical Sciences,\\
University of Durham, Durham, DH1 3LE, UK\\
$^b$Centre for Particle Theory 
and Institute for Particle Physics Phenomenology,\\
Department of Physics, University of Durham,
Durham, DH1 3LE, UK \\
E-mail: {\tt chong-sun.chu, valya.khoze, 
gabriele.travaglini@durham.ac.uk }}

\abstract{We propose a way to introduce matter fields transforming in 
arbitrary representations of the gauge group in noncommutative 
$U(N)$ gauge theories. We then argue that in the presence of supersymmetry, 
an ordinary commutative $SU(N)$ gauge theory with a general matter content 
can always be embedded, at least as an effective theory, 
into a noncommutative $U(N)$ theory at energies
above the noncommutativity mass scale $M_{\sst NC}\sim \theta^{-1/2}$.
At energies below $M_{\sst NC}$, the $U(1)$ degrees of freedom decouple 
due to the IR/UV mixing, and the noncommutative theory reduces to its
commutative counterpart. Supersymmetry can be spontaneously broken by a
Fayet-Iliopoulos D-term introduced in the noncommutative $U(N)$ theory.
$U(1)$ degrees of freedom  become arbitrarily weakly coupled in the infrared
and naturally play the r{\^o}le of the hidden sector for supersymmetry
breaking.
To illustrate these ideas we construct an effective noncommutative $U(5)$ GUT
model with Fayet-Iliopoulos supersymmetry breaking, which reduces to a 
realistic commutative theory in the infrared.
}

\keywords{Non-Commutative Geometry, Supersymmetry Breaking, Gauge Symmetry}

\preprint{{\tt hep-th/0112139}}

\begin{document}

The attempts to construct realistic models of particle physics in the framework
of noncommutative gauge field theories encounters a few outstanding problems.
First, the mixing between the infrared (IR) and the ultraviolet (UV) degrees of
freedom discovered in \cite{Minwalla} leads to novel singularities in the
low-energy effective action, further analysed in 
\cite{Matusis,Armoni,KT,Zanon,HKT,CKT,Raamsdonk,AL} (for recent reviews and a
more extensive list of references see \cite{ND,Szabo}). Second, there is an
apparent obstruction  \cite{Ter,GBM} in constructing representations of matter
fields other than (anti)-fundamental and adjoint. There is also
a problem with non-cancellation of gauge anomalies in noncommutative
theories with chiral matter content, see for example 
\cite{GBM,anom}.
In this letter we show how to
construct matter fields transforming in arbitrary representations of the
noncommutative $U(N)$ gauge group. We also explain that in supersymmetric
theories the IR/UV mixing does not lead to problems, and moreover becomes an
important ingredient of realistic model building. The problem of
anomaly cancellation in chiral theories will not be addressed in this letter.
Note however that 
our construction of general representations 
can always be applied to theories with non-chiral matter
content where the left- and right- handed fields come 
in mirror pairs. Such models are anomaly free. 

{\bf 1.} It is well-known that in non-supersymmetric theories 
the IR/UV mixing leads to IR poles in the effective action. 
This leads to drastic modifications of the dispersion 
relations for the photon \cite{Matusis} and to 
large Lorentz-violating effects \cite{ABDG} and also
threatens the renormalizability of the theory when these poles are
included at the higher loop level.
Importantly, however, this generic picture does not apply 
to supersymmetric theories, where the IR/UV mixing is 
at most logarithmic, and does not lead to either unconventional 
dispersion relations or large Lorentz-violating effects.
Moreover this milder form of IR/UV mixing leads to the 
decoupling of the $U(1)$ degrees of freedom in the IR.  
Not only the $U(1)$ degrees of freedom become free in the IR \cite{HKT}, 
they also trigger spontaneous supersymmetry breaking \cite{CKT}
in the presence of an appropriate Fayet-Iliopoulos D-term
and play the r{\^o}le of the hidden sector.

The leading order terms in the derivative expansion of the Wilsonian 
effective action for supersymmetric noncommutative QCD were analysed
in \cite{CKT}. To illustrate the decoupling of the $U(1)$ sector, it
is sufficient to look at the supersymmetric pure $U(N)$ gauge theory:
\EQ{
\cL_{\rm eff}  \ = \ 
-{1  \over 4g^{2}_{\sst 1}(k) } 
 \   F^{\sst U(1)}_{\mu \nu}   
F^{\sst U(1)}_{\mu \nu}    
\ - \ {1  \over 4g^{2}_{\sst N}(k) } 
 \   F^{\sst SU(N)}_{\mu \nu}   
F^{\sst SU(N)}_{\mu \nu} 
\ + \ 
\cdots
\ , 
\label{uoneres}}
where the dots stand for terms involving fermions and 
higher-derivative corrections. 
The multiplicative coefficients in front of the gauge kinetic terms in
\eqref{uoneres} define the Wilsonian coupling constants. 
The running of the $U(1)$ 
has the following asymptotic behaviour:
\EQ{
{1\over g^{2}_{\sst 1}(k)}  \rightarrow
\pm {{3N} \over (4\pi)^2} \log k^2 \ , 
\label{r1}}
where the plus (minus) sign corresponds to $k^2\to\infty$ 
($k^2\to 0$), whereas for the $SU(N)$ gauge factor we have
in both limits:
\EQ{
{1\over g^{2}_{\sst N}(k)}  \rightarrow
{{3N} \over (4\pi)^2} \log k^2 \ . }
The change in the running of the $U(1)$ coupling in \eqref{r1}
occurs at the scale
$k^2 \sim M_{\sst NC}^2,$ where $M_{\sst NC}\sim \theta^{-1/2}$ is the
noncommutative mass and $\theta$ the usual noncommutativity parameter.
This running was interpreted in \cite{CKT} as having a full noncommutative
$U(N)$ gauge theory in the UV, which in the low-energy limit appears as a commutative $SU(N)$ theory with a decoupled free $U(1)$ factor.
Note that this decoupling in the IR does not mean that the noncommutative
$U(N)$ gauge symmetry is broken. In fact, there is a gauge-invariant completion
of \eqref{uoneres} proposed in \cite{Raamsdonk,AL} which involves open
Wilson lines \cite{IIKK}. This completion of \eqref{uoneres} introduces 
higher-derivative terms which are irrelevant for the low-energy dynamics.

Finally, note that
supersymmetry can be spontaneously broken by introducing a Fayet-Iliopoulos
D-term in the lagrangian
\EQ{\cL_{\sst FI} =\, \xi_{\sst FI} \int  d^2 \theta d^2 \thetabar
\; \tr_{\sst N} V
\ ,
\label{fiterm}
}
where $V$ is the real $U(N)$ vector superfield and the trace over the
$N$ by $N$ matrices selects the $U(1)$-component of $V$.
The Fayet-Iliopoulos 
action, $\int d^4 x L_{\sst FI},$ is $U(N)$ gauge invariant and can be
naturally introduced at tree-level in any $U(N)$ theory.
This provides us with a scenario of a gauge-mediated supersymmetry breaking
where the $U(1)$ degrees of freedom, which 
eventually become arbitrarily weakly coupled in the IR,
play the r{\^o}le of the hidden sector \cite{CKT}. Both the hidden sector and the messenger sector are naturally part of the noncommutative $U(N)$ gauge theory.

{\bf 2.} Now we discuss how to introduce matter fields transforming in 
general representations of noncommutative $U(N)$. The construction of 
fundamental, $f^i(x),$ anti-fundamental, $\ft_i(x),$  and adjoint,
$\phi^i_j(x)$ representations of $U(N)$ is straightforward,
\EQ{f^i \to  U^i_{i'} * f^{i'} \ , \quad 
\ft_i \to \ft_{i'} * (U^{-1})^{i'}_i \ , \quad
\phi^i_j \to U^i_{i'} *\phi^{i'}_{j'} * (U^{-1})^{j'}_j \ ,
}
where the $*$-product is the usual Weyl-Moyal deformation,  
$(f * g) (x) \equiv f(x) e^{{i\over 2}\theta^{\mu\nu}
\stackrel{\leftarrow}{\partial_\mu}
\stackrel{\rightarrow}{\partial_\nu}}  g(x),$
and $U\in U(N)$.
Consider now other representations, for example, a
rank two representation $t^{ij}(x)$. 
The naive noncommutative gauge transformation, 
$t^{ij} \to U^i_{i'} * U^j_{j'} *t^{i'j'},$
is not correct as the closure property of the group multiplication,
\EQ{
(t^U)^V = t^{U*V} \ ,
\label{closure} 
}
is broken due to the noncommutativity of the $*$-product. 
It is easy to convince oneself
that rearranging the positions of the $U$'s or any other straightforward
modification of the naive transformation does not help. This
is in fact a generic problem well-known in noncommutative geometry.  

To resolve this problem  we propose to
modify the transformation laws for the  
matter fields in a non-trivial, gauge field-dependent way. To this end
we first introduce gauge-singlet matter fields,
$\cT^{ij}(x),$ as in the construction
of noncommutative baryons of Ref.~\cite{CD},
making use of the following open Wilson line 
\EQ{
W(x) = P_{*} \exp \left( i \int_{0}^{1} d\sigma \ 
{d\zeta^{\mu}\over d \sigma}
A_\mu (x + \zeta (\sigma) ) \right) \ , \label{wline}
}
where the integration is along the contour $C_{\infty}$ from $\infty$ to $x,$
\EQ{
C_{\infty} = \{ \zeta_{\mu} (\sigma ), \ 0\leq \sigma \leq 1\ | \ 
\zeta (0) = \infty, \ \zeta (1) = 0 \} \ , \label{ctr}
}
and the path ordering is with respect to the star product. 
The shape of the contour is not important for our construction, but for concreteness one can always consider straight rays such as
the one from $(y_0,y_1,y_2,\infty)$ to $(x_0,x_1,x_2,x_3).$
Under a noncommutative gauge transformation,
$A_\mu \to U*(A_\mu -i \partial_\mu)*U^{-1},$
the Wilson line \eqref{wline} transforms as
\EQ{W(x) \to U_\infty * W(x) * U^{-1}(x) \ . 
\label{wtranold}}
The key element of our construction is to restrict the allowed gauge
transformations $U(x)$ to those which, as $x\to \infty$ 
{\it along the contour} \eqref{ctr}, approach
a constant $U_{\infty},$
with vanishing derivatives to all orders.
Then, without loss of generality we can set $U_\infty = 1,$ and
as a result \eqref{wtranold} becomes 
\EQ{W(x) \to W(x) * U^{-1}(x) \ .
\label{wtrannew} }
This restriction can be motivated in a number of ways. For example,
it is compulsory when the theory is compactified on a 4-sphere.
Also the Noether charges associated to the gauge symmetry transform
covariantly (form an algebra) only under gauge transformations
which approach the identity at spatial infinity
\cite{Jackiw}. In addition,
if we first fix the $A_0=0$ gauge, the requirement of 
$U(|\vec{x}|=\infty) = 1$ is necessary to project onto the states
which satisfy the Gauss' law \cite{Rossi:1980jf,GPY}.

We can now associate to $t^{ij}$ a gauge-singlet field $\cT^{ij}$ 
defined by 
\EQ{
\cT^{ij} = W^{i}_{i'} * W^{j}_{j'} *t^{i'j'}
\ .
}
The invariance of $\cT^{ij}$ determines the transformations 
of $t^{ij}$ under the noncommutative gauge group, 
\EQ{
(t^U)^{ij} = (U*W^{-1})^{j}_{k} * U^{i}_{l}  *W^{k}_{m} *
t^{l m} \label{tU}
\ .
}
In tensor notation, it reads
\EQ{
t^U = (1 \otimes U*W^{-1}) * (U \otimes W ) * t
\ .
\label{ttrans}
}
Remarkably, this transformation satisfies
the closure property of group multiplication \eqref{closure}.
In the commutative limit the Wilson lines in \eqref{ttrans}
will cancel each other and the transformation law will reduce
to the conventional one.

The same construction applies to all higher-rank representations.
For a generic rank-$n$ representation $t_{[n]}$, the generalization of 
\eqref{ttrans} reads
\EQ{t_{[n]}^{U} =  (U * W^{-1})^{\e_n}_{n} *
(U * W^{-1})^{\e_{n-1}}_{n-1}* \cdots * (U * W^{-1})^{\e_1}_1 *
(W^{\e_1} \otimes \cdots \otimes W^{\e_n} ) * t_{[n]}
\ . \label{ttt}}
Here we have defined 
\EQ{
(U * W^{-1})^{\e_i}_{i} \equiv 
1 \otimes \cdots \otimes (U * W^{-1})^{\e_i} \otimes \cdots \otimes 1
\ ,
}
with $(U * W^{-1})^{\e_i}$ in the $i^{\rm th}$ position in the tensor product,
and the power $\e_i = +1\,  (-1)$ if $i$ is 
an upper (lower) index.
Equation \eqref{ttt} follows from the invariance of the rank-$n$ 
gauge-singlet field
\EQ{\cT_{[n]}=(W^{\e_1}\otimes \cdots \otimes W^{\e_n}) * t_{[n]} \ .}
Irreducible representations are obtained from the reducible ones 
in the same way as in the commutative case.

Remarkably, the gauge-singlet matter fields $\cT(x)$ introduced
above are related to $t(x)$ by a gauge transformation $U=W,$
\EQ{ \cT=t^{W} \ .\label{16}}
By the same token we now introduce the gauge-singlet vector field
\EQ{ {\cal A}_\mu \equiv A_\mu^{W}=W * (A_\mu-i\partial_\mu)*W^{-1} \ ,
\label{caaw}}
and write down the appropriate gauge-singlet `covariant' derivative for
matter fields. In the rank 2 case we have
\EQ{ \cD_\mu = (1\otimes 1) \partial_\mu + i (
{\cal A}_\mu \otimes 1) 
+ i (1 \otimes{\cal A}_\mu) \ ,
}
and the generalization to the rank-$n$ case is obvious.
With these ingredients we can now construct a gauge-invariant action
for the matter field $t$ 
\beqa
\int d^4x\, \tr \,|\cD_\mu * t^W|^2 &=& \int d^4x\, \tr\, |D_\mu *t|^2 \ , \quad
\ {\rm for\ scalars}, \\
i\int d^4x\, \tr\, \bar{t}^W *\gamma^\mu \cD_\mu *t^W &=&
i\int d^4x\, \tr\, \bar{t}* \gamma^\mu D_\mu *t \ , \quad
{\rm for\ fermions},
\eeqa
where $D_\mu$ is defined such  that $\cD_\mu * t^W=(D_\mu *t)^W.$
It then follows that
\EQ{D_\mu = (W\otimes W)^{-1} *\left(
(1\otimes 1) \partial_\mu + i (
{A}^W_\mu \otimes 1) 
+ i (1 \otimes{A}^W_\mu)
\right)* (W\otimes W) \ ,
}
and ${A}^W_\mu$ was defined in \eqref{caaw}.
The action written in terms of the original variables $t(x)$ and $A_\mu(x)$
is gauge-invariant, but takes a cumbersome form which is not tractable
in perturbation theory: Taylor-expanding the Wilson lines would lead
to non-renormalizable vertices and is not a good idea as it misses the
fact that $W(x)$ is a $U(N)$ group element. For practical applications, 
one first has to perform a gauge transformation using
$U(x)=W(x)$ on the original variables $t(x)$ and $A_\mu(x)$, arriving
at the gauge-singlet variables 
as in \eqref{16}, \eqref{caaw}. The action takes now exactly the same
form as in the commutative theory (with star products). This transition
to the gauge-singlet variables is nothing but a gauge-fixing procedure.
For example, we can choose $W(x)$ to be 
a straight Wilson line parallel to the $x_3$ axis.
This Wilson line is precisely the gauge transformation used to fix the
$A_3=0$ gauge. Moreover, the usual
residual  $x_3$-independent gauge transformations are not allowed since
$U_\infty=1$, the gauge fixing is complete and no further gauge
transformations are possible. Hence we can interpret the gauge-singlet
fields as the degrees of freedom of the completely gauge-fixed formulation.
Wilson lines and non-renormalizable interactions are absent 
in this `physical' gauge which is suitable for perturbative calculations.

A few remarks are in order. First,
in the ordinary commutative case one can still carry out this construction,
which however trivially reduces to the usual approach with standard
gauge transformations, since the Wilson lines would cancel in  \eqref{ttrans}
and \eqref{ttt} 
as remarked earlier. Second, the same comment applies also to the
noncommutative theory with (anti)-fundamental and adjoint matter fields.
Third, the supersymmetrization of this construction is immediate
in terms of component fields. To establish a gauge-invariant formulation 
in the superfield formalism, one can similarly
introduce gauge-singlet vector superfields $\cV(x,\vartheta,\bar{\vartheta})$
(at least in the Wess-Zumino gauge) 
\EQ{e^{2\cV} = (W^{-1})^\dagger (\bar{y})* e^{2V(x,\vartheta,\bar{\vartheta})}*
W^{-1}(y) \label{vsfsin}
\ , } 
where $y$ is the usual chiral coordinate in the superspace and the 
singlet chiral matter superfields $\cT(y,\vartheta)$ in the rank 2
case are given by
\EQ{ \cT(y,\vartheta)= (W (y)\otimes W (y))* t(y,\vartheta) \label{csfsin}
\ . } 
Then the action $S[\cV,\cT]$ takes the standard form
$\int d^4 x \int d^2 \vartheta d^2 \bar{\vartheta}
\,\cT^\dagger * ( e^{2\cV}\otimes e^{2\cV}) * \cT.$

Finally, note that we can give any singlet matter field  an
arbitrary $U(1)$ charge via the coupling to the gauge kernel
$K=\,\exp\, [2\cV -{\rm const}\, \tr_N \cV].$ In doing so one may
lose the renormalizability of the noncommutative $U(N)$ theory,
and such models should be thought as effective field theories.
In the limit $\theta_{\mu\nu}\to 0$ these models
reproduce the commutative theories with product gauge group
$U(1)\times SU(N).$ 
In the presence of supersymmetry, the IR/UV mixing amounts to nothing more 
than the decoupling in the low-energy limit
of the overall $U(1)$ factor, as explained in {\bf 1.} 
Using our construction,
an ordinary commutative $U(1)\times SU(N)$ gauge theory 
with matter fields in general representations 
can always be embedded into a supersymmetric
noncommutative $U(N)$ theory at energies
above the noncommutativity mass scale $M_{\sst NC}$.

{\bf 3.} To illustrate the machinery we have discussed, we now consider
a simple supersymmetric noncommutative unified theory (NUT) with gauge group $U(5)$, with an adjoint Higgs superfield $\Phi$ 
and three generations of matter superfields $\tilde{\cF}$ and $\cT$ 
in the anti-fundamental and in the rank 2 antisymmetric representations generalizing \cite{gut}. This is a chiral theory with nonvanishing
gauge anomalies and it should be taken as an effective field theory.

We work in the physical gauge introduced in {\bf 2}.
The matter fields are coupled to the gauge fields as follows:
\EQ{\int d^4 x \int d^2 \vartheta d^2 \bar{\vartheta}
\left(\tilde{\cF} * K_{\tilde{\cF}} *\tilde{\cF}^\dagger\, +\,
\cT^\dagger * (K_{\cT} \otimes K_{\cT}) * \cT
\right) \ , }
where the gauge kernels are
\EQ{
K_{\tilde{\cF}} = e^{-2\left(\cV-\uno\, 
{\kappa_{\tilde{\cF}}\over N}\tr_N \cV \right)}
\ ,
\quad 
K_{\cT} = 
e^{2\left(\cV-\uno\, 
{\kappa_{\tilde{\cT}}\over N}\tr_N \cV \right)}
 \ ,
} 
and  $\kappa_{\tilde{\cF}}$ and $\kappa_{\cT}$ are arbitrary constants.
Furthermore, to break supersymmetry we introduce the
Fayet-Iliopoulos term \eqref{fiterm}.

Below the noncommutativity mass scale the IR/UV mixing triggers the
decoupling of the overall $U(1)$ degrees of freedom from the $SU(5)$
as explained in {\bf 1.} Then the $SU(5)$ gauge symmetry gets broken
by the Higgs vacuum expectation value
\EQ{
\langle \phi \rangle = M_{\sst NUT}\, {\rm diag} (1,1,1,-3/2,-3/2)
}
below the NUT scale $M_{\sst NUT}<M_{\sst NC}$
to $SU(3)\times SU(2) \times U(1).$ The matter fields decompose into
representations of $SU(3)\times SU(2)$ as 
${\bf \bar{5}}=({\bf \bar{3}},{\bf {1}}) + ({\bf {1}},{\bf {2}})$ and
${\bf {10}}=({\bf {3}},{\bf {2}}) +({\bf \bar{3}},{\bf {1}}) + 
({\bf {1}},{\bf {1}}).$

We can now solve the D-flatness conditions:
\beqa
D^0&=&\sqrt{2/ N}
( 1- \k_{\cT})\, \cT^\dagger  \cT - 
\sqrt{2/ N}(1-\k_{\tilde{\cF}}) 
\,\tilde{\cF} \tilde{\cF}^\dagger -\xi_{\sst FI},\\
D^a &=& 2\cT^\dagger (1\otimes T^a) \cT - 2 \tilde{\cF} T^a \tilde{\cF}^\dagger
\ , \quad a=1,\ldots, N^2-1\ , \ (N=5) \ .
\eeqa
 In general, $\langle D^a\rangle =0$ and $\langle D^0\rangle =-\xi_{\sst FI}$
and supersymmetry
is broken spontaneously for $\xi_{\sst FI}\neq 0$.
For the choice   
$\kappa_{\tilde{\cF}}=0$ and $\kappa_{\cT}=2$
we get a
phenomenologically interesting supersymmetry breaking mass spectrum,
in which the masses for all the matter scalar fields (squarks and sleptons)
are shifted by a positive amount $\sim \sqrt{\xi_{\sst FI}}.$
We make no claim, however, that this simple model is realistic
or should be taken as a microscopic quantum field theory.
For the reasons of anomaly cancellation, nonchiral noncommutative
theories at present look more realistic.

\section*{Acknowledgements} 
We would like to thank Giovanni Amelino-Camelia, Harald Dorn,
Gordy Kane and Tim Morris for discussions.
This work was partially supported by a PPARC SPG grant.

\ed